\documentclass[cits]{PoS}
\usepackage{amsmath}

\newcommand{\Tr}{\mbox{Tr}}

\def\gtwid{{\,\raise.3ex\hbox{$>$\kern-.75em\lower1ex\hbox{$\sim$}}\,}}
\def\ltwid{{\,\raise.3ex\hbox{$<$\kern-.75em\lower1ex\hbox{$\sim$}}\,}}

\def\chpt{\raise0.4ex\hbox{$\chi$}PT}
\def\schpt{S\raise0.4ex\hbox{$\chi$}PT}
\def\rschpt{rS\raise0.4ex\hbox{$\chi$}PT}
\def\wchpt{W\raise0.4ex\hbox{$\chi$}PT}

\title{Wilson chiral perturbation theory, Wilson-Dirac operator eigenvalues
and clover improvement\footnote{To appear as PoS (Confinement X) 077}}

\ShortTitle{Wilson chiral perturbation theory, Wilson-Dirac operator eigenvalues \dots}

\author{Poul H. Damgaard\\
        Niels Bohr International Academy and Discovery Center, Niels Bohr Institute\\
        University of Copenhagen\\
        Blegdamsvej 17, DK-2100 Copenhagen \O, Denmark}

\author{\speaker{Urs M. Heller}%
	\\
        American Physical Society, One Research Road, Ridge, NY 11961, USA\\
        E-mail: \email{heller@aps.org}}

\author{Kim Splittorff\\
        Discovery Center, Niels Bohr Institute, University of Copenhagen\\
        Blegdamsvej 17, DK-2100 Copenhagen \O, Denmark}

\abstract{
Chiral perturbation theory for eigenvalue distributions, and equivalently
random matrix theory, has recently been extended to include lattice effects
for Wilson fermions. We test the predictions by comparison to eigenvalue
distributions of the Hermitian Wilson-Dirac operator from pure gauge
(quenched) ensembles. We show that the lattice effects are diminished
when using clover improvement for the Dirac operator. We
demonstrate that the leading Wilson low-energy constants associated with
Wilson (clover) fermions can be determined using spectral information of
the respective Dirac operator at finite volume.
}

\FullConference{Xth Quark Confinement and the Hadron Spectrum\\
		October 8 - 12, 2012\\
		TUM Campus Garching, Munich, Germany}

\begin{document}

\section{Introduction}

The low-energy behavior of QCD, the spontaneous breaking of chiral symmetry,
including the explicit breaking by the quark masses, is described by
chiral perturbation theory (\chpt). In the lattice regularization of QCD
lattice artifacts can contribute to the breaking of chiral
symmetry, for Wilson fermions, or its partial breaking, in the case of
staggered fermions. These effects can be included in the \chpt\ approach,
leading to new, lattice discretization dependent low-energy constants.
We consider Wilson fermions in this contribution, for which the effective
theory, Wilson \chpt\ (\wchpt) was introduced and worked out in 
\cite{WCHiPT}. The new terms in the chiral Lagrangian affect the low-lying
spectrum of the (Hermitian) Wilson-Dirac operator \cite{WRMT}.
For a recent review with additional references, see Ref.~\cite{WCHiPT_rev}.
Here, we test and verify the predictions for the distribution of the low-lying
eigenvalues with lattice QCD simulations \cite{WRMT_num,WRMT_num2}
and show \cite{WRMT_num3} that they can be used to obtain the new
low-energy constants introduced in \wchpt. We also demonstrate the effect
of clover improving the Wilson-Dirac operator.

\section{The \wchpt\ and Wilson RMT framework}

We will be concerned with the $\epsilon$-regime of \wchpt\ where the zero momentum modes dominate -- the system size is such that $m_\pi L \ll 1$.
In addition we adopt the power counting with $m \sim a^2$. Hence, dropping
the kinetic part of the chiral Lagrangian, we consider
\begin{equation}
{\cal L} =  -\frac{1}{2} m \Sigma \Tr \left( U + U^\dagger \right)
 -\frac{1}{2} z \Sigma \Tr \left( U - U^\dagger \right) + a^2 {\cal V} ~.
\label{eq:L_chpt}
\end{equation}
The second term, representing a $\bar\psi \gamma_5 \psi$ term, is
introduced for later convenience.
${\cal V}$ describes the lattice artifacts \cite{WCHiPT}
\begin{equation}
{\cal V} = W_8 \Tr \left( U^2 + U^{\dagger 2} \right)
 + W_6 \left[ \Tr \left( U + U^\dagger \right) \right]^2
 + W_7 \left[ \Tr \left( U - U^\dagger \right) \right]^2 ~.
\label{eq:W_chpt}
\end{equation}
At large $N_c$, the two-trace terms are suppressed.

The finite size scaling considered is such that
\begin{equation*}
\hat{m} = m \Sigma V ~, \qquad \hat{z} = z \Sigma V \qquad \text{and} \qquad
\hat{a}_j^2 = a^2 W_j V \quad \text{for} \quad j = 6, 7, 8
\end{equation*}
are held fixed. Here $\Sigma$ is the condensate and $V$ the volume.

This leading order in \wchpt\ can equivalently be
described by a chiral random matrix theory (RMT). For Wilson fermions,
including the one-trace term with low-energy constant $W_8$, the Dirac
operator is represented in Wilson RMT (WRMT) as \cite{WRMT}
\begin{equation}
{\cal D}_W = \left( \begin{array}{cc}
\tilde{a} A & i W \\ i W^\dagger & \tilde{a} B
\end{array} \right) ~,
\label{eq:D_WRMT}
\end{equation}
with $W$ a random $(n+\nu) \times n$ complex matrix, and $A$ and $B$ random
Hermitian matrices of size $(n+\nu) \times (n+\nu)$ and $n \times n$,
respectively. As usual in the RMT context, we consider
a fixed index $\nu$. We use a chiral basis with
$\gamma_5 = \text{diag}(1, \dots, 1, -1, \dots, -1)$.
$A$ and $B$ represent the chiral symmetry breaking term
corresponding to the Wilson term in the Wilson-Dirac operator.

The two-trace terms can be incorporated in WRMT via two Gaussian integrations
\begin{equation}
Z^\nu(\hat{m}, \hat{z}; \hat{a}_6, \hat{a}_7, \hat{a}_8) =
 \frac{1}{16\pi \hat{a}_6 \hat{a}_7} \int_{-\infty}^{\infty}
 dy_6 dy_7 \rm{e}^{-\frac{y_6^2}{16 \hat{a}_6^2} -\frac{y_7^2}{16 \hat{a}_7^2}}
 \, Z^\nu(\hat{m} - y_6, \hat{z} - y_7; 0, 0, \hat{a}_8) ~.
\label{eq:Zfull_WRMT}
\end{equation}
Here,
\begin{equation}
Z^\nu(\hat{m}, \hat{z}; 0, 0, \hat{a}_8) =
\int dU \det{}^\nu U \rm{e}^{- V {\cal L}(W_6=W_7=0) }
\label{eq:Z_WRMT}
\end{equation}
is the fixed-index partition function with the one-trace ${\cal O}(a^2)$
term included.

\section{Index of the Wilson-Dirac operator}

As indicated above, RMT predictions apply to gauge field sectors with
a fixed index, or, in the continuum, fixed topological charge. For the
Wilson-Dirac operator, the index can be defined by
\begin{equation}
\nu ~\equiv~ {\sum_{k}}' {\rm sign} (\langle k|\gamma_5|k\rangle)
\label{eq:index}
\end{equation}
with $|k\rangle$ the $k$'th eigenstate of the Wilson-Dirac operator, $D_W$.
Only eigenvectors with real eigenvalues contribute, and the $\prime$
indicates that only the real eigenvalues in the branch near zero,
with eigenvalues $< r_{cut}$, are kept. Introducing the Hermitian
Wilson-Dirac operator $D_5(m_0) = \gamma_5 (D_W + m_0)$ and using
\begin{equation}
D_5(m_0) |\psi\rangle = 0 \qquad \Rightarrow \qquad
D_W |\psi\rangle = -m_0 |\psi\rangle
\label{eq:realEv}
\end{equation}
the index can equivalently be obtained from the zero crossings of the
spectral flow of $D_5(m_0)$ up to $m_{cut} = - r_{cut}$ \cite{spec_flow}.
It corresponds to the index of an overlap operator \cite{overlap} with
kernel $D_5(m_{cut})$.  Because of the dependence on the choice of
$r_{cut}$, the index of the Wilson-Dirac operator is not unique.

\section{The numerical simulations}

For our numerical tests, in the quenched case, we generated three
ensembles using the Iwasaki gauge action \cite{Iwasaki}, which suppresses
dislocations and gives a fairly unique index $\nu$ or topological
charge $Q$. The ensembles are characterized in Table~\ref{tab:Ensembles}.

\begin{table}
\begin{center}
\begin{tabular}{|c|c|c|c|c|c|c|}
\hline
Ens & $\beta_{Iw}$ & $r_0/a$ & $a$ [fm] & size & $L$ [fm] &
 $|Q| =$ 0, 1, 2 cfgs \\
\hline
A & 2.635 & 5.37 & 0.093 & $16^4$ & 1.5 & 1279, 2257, 1530 \\
B & 2.635 & 5.37 & 0.093 & $20^4$ & 1.9 & \phantom{1}401, \phantom{1}682,
 \phantom{1}644 \\
C & 2.79  & 6.70 & 0.075 & $20^4$ & 1.5 & 1207, 2130, 1448 \\
\hline
\end{tabular}
\end{center}
\caption{The ensembles used. The scale is set by $r_0 = 0.5$ fm. The $r_0/a$
values come from interpolation formulae in \protect\cite{Iw_r0}. $Q$ is
the topological charge (see text).}
\label{tab:Ensembles}
\end{table}

The topological charge listed in Table~\ref{tab:Ensembles} was obtained
after six steps of HYP smearing \cite{HYP} with an improved lattice
$F \tilde{F}$ operator \cite{Boulder_Q}. On the configurations with
$|Q| \le 1$, as well as the $|Q|=2$ configurations of ensemble A, we
also did the much more expensive computation of the index from the
spectral flow. We first applied one HYP smearing before constructing
the Wilson-Dirac operator. The topological charge and the index agreed
on most configurations, with the agreement improving at smaller lattice
spacing and becoming worse for the larger volume ensemble C, for which
it was about 97\%.

\begin{figure}
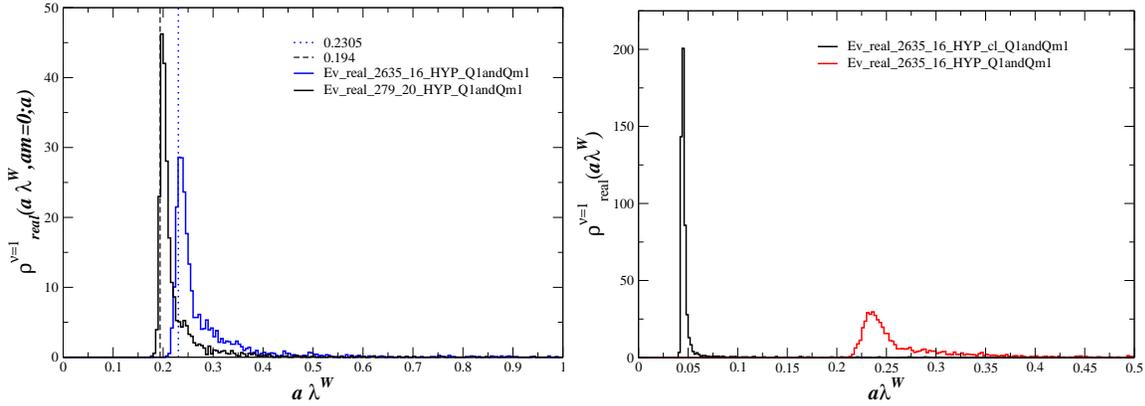

\begin{center}
\begin{tabular}{c c}
\hspace{-0.4truecm}
\includegraphics[width=7.45cm]{Ev_real_2635_16_HYP_Q1VS279_20_HYP_Q1-rhoreal-lattice-units.eps}
&
\hspace{-0.4truecm}
\includegraphics[width=7.55cm]{Ev_real_2635_16_HYP_Q1andQm1_no-clVScl.eps}
\end{tabular}
\end{center}
\vspace{-0.6truecm}
\caption{Distribution of the real eigenvalues of the Wilson-Dirac operator
for the $\nu=1$ configurations of the two $L=1.5$ fm ensembles A and C (left)
and comparison of the distribution with and without clover improvement at
$a=0.093$ fm (ensemble A, right).}
\label{fig:real_Ev}
\end{figure}

The crossing points in the spectral flow are the real eigenvalues,
whose distribution is shown in Fig.~\ref{fig:real_Ev} (left). The
dashed vertical lines are estimates of (minus) the critical mass.
Some real eigenvalues are smaller, on so-called ``exceptional''
configurations.

For ensemble A, we also computed the spectral flow with clover improving
the Wilson-Dirac operator, again after one HYP smearing. The clover
coefficient was set to 1, which is expected to be close to the
nonperturbative value after the HYP smearing \cite{csw_NP}. The resulting
distribution of the real eigenvalues is compared to the unimproved case in
Fig.~\ref{fig:real_Ev} (right). The improvement is quite dramatic, besides
the expected reduced shift away from zero, the distribution is much
narrower and more symmetric. The width is determined by the ${\cal O}(a^2)$
terms in Eq.~(\ref{eq:W_chpt}), so with clover improvement the coefficients
are much smaller.

\section{Wilson eigenvalue distributions and WRMT}

We next computed the lowest 20, in magnitude, eigenvalues of the Hermitian
Wilson-Dirac operator $D_5(m_0)$ with bare mass $am_0=-0.216$ for ensembles
A and B to compare to eigenvalues distributions obtained from WRMT
\cite{WRMT}.

\begin{figure}[h]
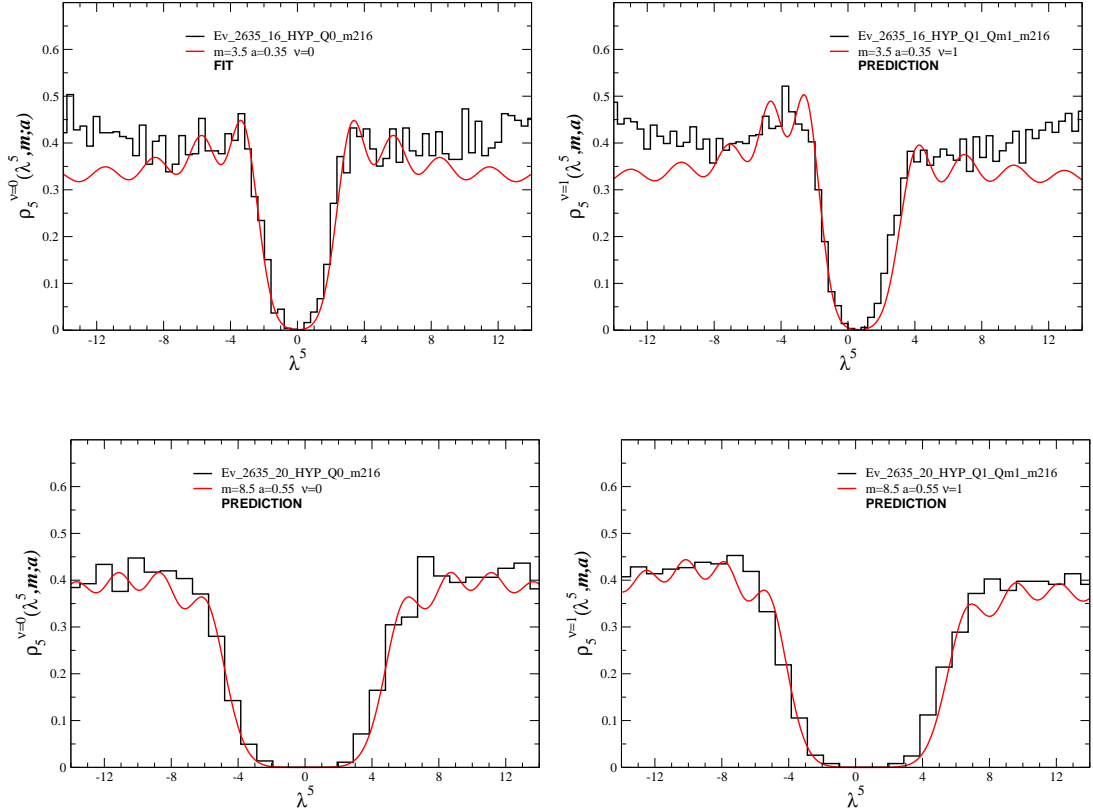

\begin{center}
\vspace{0.1truecm}
\begin{tabular}{c c}
\hspace{-0.4truecm}
\includegraphics[width=7.0cm]{Ev_2635_16and20_HYP_Q0_m216-16.eps}
&
\hspace{-0.2truecm}
\includegraphics[width=7.0cm]{Ev_2635_16and20_HYP_Q1_Qm1_m216-16.eps}
\end{tabular}
\vskip 0.7cm
\begin{tabular}{c c}
\hspace{-0.2truecm}
\includegraphics[width=7.0cm]{Ev_2635_16and20_HYP_Q0_m216-20.eps}
&
\hspace{-0.2truecm}
\includegraphics[width=7.0cm]{Ev_2635_16and20_HYP_Q1_Qm1_m216-20.eps}
\end{tabular}
\end{center}
\vspace{-0.6truecm}
\caption{Comparison of the histogrammed eigenvalue distributions with
WRMT. The $\nu=0$ distribution of ensemble A (top left) was used to
obtain the parameters. The predictions for ensemble B (bottom) used
volume scaling of the parameters.}
\label{fig:Ev_Wils_2635}
\end{figure}

Without clover improvement, we considered only contributions from the
two-trace term in Eq.~(\ref{eq:W_chpt}) \cite{WRMT_num}.  We used the
$\nu=0$ histogrammed eigenvalue distributions of ensemble A to determine
the WRMT parameters $\hat{m}$ and $\hat{a}=\hat{a_8}$ and the eigenvalue
rescaling factor $\Sigma V$. Using the same parameters we then get a
prediction for the $\nu=1$ distribution that can be compared to the
numerical data (see Fig.~\ref{fig:Ev_Wils_2635} top).

Ensemble B differs from ensemble A only in the volume. Using volume
scaling, $\hat{m}_B = \hat{m}_A (V_B/V_A)$ and $\hat{a}_B = \hat{a}_A
\sqrt{V_B/V_A}$, we obtain predictions for the distributions for
ensemble B (see Fig.~\ref{fig:Ev_Wils_2635} bottom). As can be seen,
the WRMT predictions work well.

\begin{figure}[h]
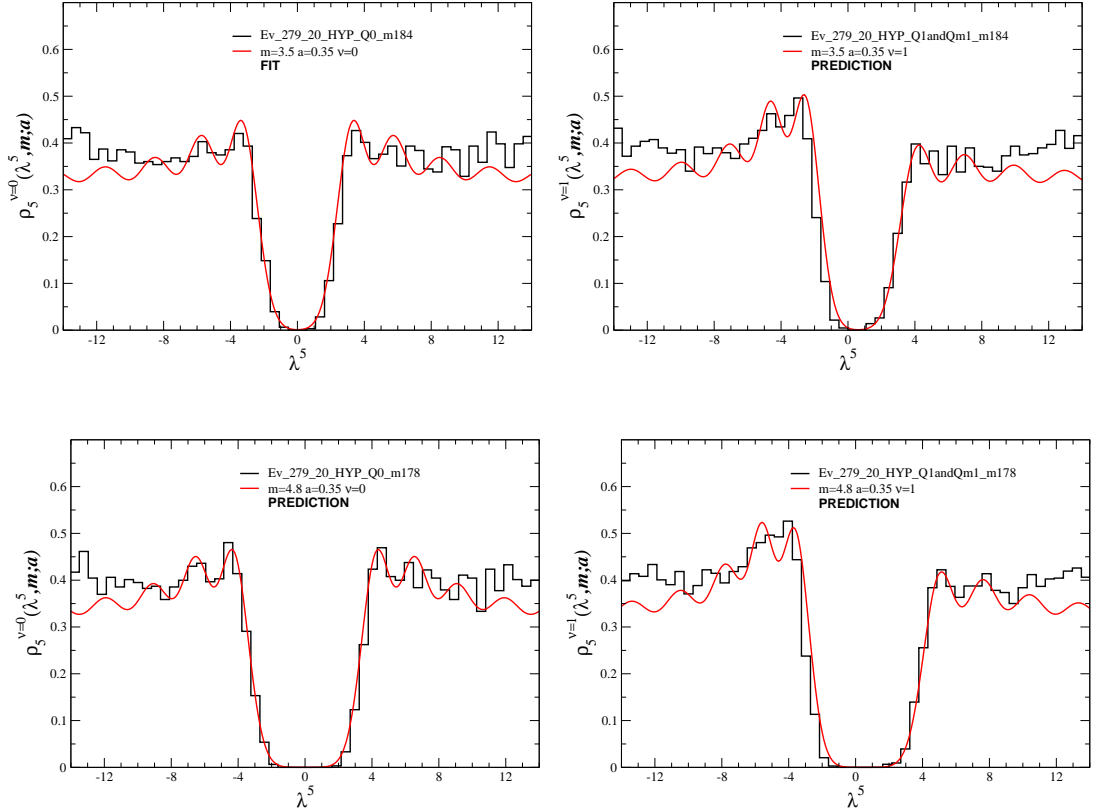

\begin{center}
\vspace{0.1truecm}
\begin{tabular}{c c}
\hspace{-0.4truecm}
\includegraphics[width=7.0cm]{Ev_279_20_HYP_Q0_m184-1172configs-fit2-v2.eps}
&
\hspace{-0.2truecm}
\includegraphics[width=7.0cm]{Ev_279_20_HYP_Q1andQm1_m184-990resp988configs-fit2-v2.eps}
\end{tabular}
\vskip 0.7cm
\begin{tabular}{c c}
\hspace{-0.2truecm}
\includegraphics[width=7.0cm]{Ev_279_20_HYP_Q0_m178-1172configs-m-scaling-v1.eps}
&
\hspace{-0.2truecm}
\includegraphics[width=7.0cm]{Ev_279_20_HYP_Q1andQm1_m178-990resp988configs-fit2-m-scaling.eps}
\end{tabular}
\end{center}
\vspace{-0.6truecm}
\caption{Comparison of the histogrammed eigenvalue distributions with
WRMT for ensemble C. The $\nu=0$ distribution with bare mass $am_0=-0.184$
(top left) was used to obtain the parameters. The predictions for bare
mass $am_0=-0.178$ (bottom) used ``mass scaling'', $\Delta \hat{m} =
\Delta m_0 \Sigma V$.}
\label{fig:Ev_Wils_279mass}
\end{figure}

For ensemble C, at the smaller lattice spacing, we computed the eigenvalues
with two different bare masses $am_0=-0.178$ and $-0.184$.  We used the
histogrammed $\nu=0$ distribution with bare mass $am_0=-0.184$ to the
determine the WRMT parameters, and used ``mass scaling'', $\Delta \hat{m} =
\Delta m_0 \Sigma V$ for predictions for the distributions with the other
bare mass $am_0=-0.178$, as shown in Fig.~\ref{fig:Ev_Wils_279mass}.
Again, the WRMT predictions work well.

\section{Clover improved eigenvalue distributions and WRMT}

\begin{figure}[h]
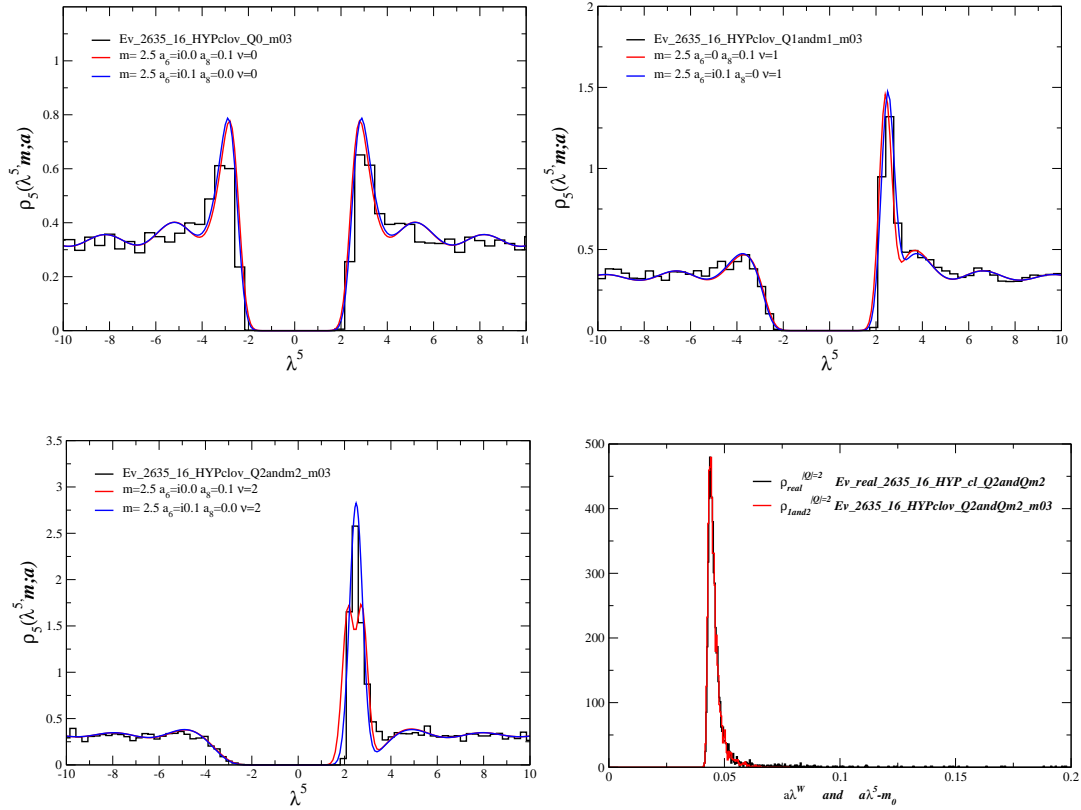

\begin{center}
\vspace{0.1truecm}
\begin{tabular}{c c}
\hspace{-0.4truecm}
\includegraphics[width=7.0cm]{Ev_2635_16_HYPclov_Q0_m03.eps}
&
\hspace{-0.4truecm}
\includegraphics[width=7.0cm]{Ev_2635_16_HYPclov_Q1andm1_m03-update.eps}
\end{tabular}
\vskip 0.65cm
\begin{tabular}{c c}
\hspace{-0.2truecm}
\includegraphics[width=7.0cm]{Ev_2635_16_HYPclov_Q2andm2_m03.eps}
&
\hspace{0.1truecm}
\includegraphics[width=6.6cm]{Ev_2635_16_HYP_cl_realVS12.eps}
\end{tabular}
\end{center}
\vspace{-0.6truecm}
\caption{Comparison of the histogrammed clover-improved eigenvalue
distributions with WRMT for ensemble A (top and bottom left). The
$\nu=1$ distribution (top right) was used to obtain the WRMT parameters.
The red curves are the WRMT predictions with $\hat{a}_8 \ne 0$, the blue
curves those with $\hat{a}_6 \ne 0$. The bottom right plot shows a
comparison of the distribution of the real eigenvalues with the first
two positive eigenvalues of $D_5(m_0)$ shifted by the bare mass for the
$\nu=2$ configurations.}
\label{fig:Ev_clov_2635}
\end{figure}

We have already seen from the distribution of the real eigenvalues in
Fig.~\ref{fig:real_Ev} (right) that clover improvement not only, as
expected, decreases the additive mass renormalization (the real eigenvalue
peak is much closer to zero) but also the size of the ${\cal O}(a^2)$
low-energy constants considerably (the distribution becomes much narrower).
Here we consider the effects on the distribution of the 20 lowest, in
magnitude, eigenvalues of the Hermitian Wilson-Dirac operator $D_5(m_0)$
with clover improvement for ensemble A using a bare mass $am_0=-0.03$. The
comparison with WRMT is shown in the first three panels of
Fig.~\ref{fig:Ev_clov_2635}.

For $|\hat{a}_j| \ll 1$ the lattice effects affect, to leading order, only
the index peak of the topological modes. These are the lowest eigenvalues
with almost chiral eigenvectors which correspond to the real eigenvalues
shifted by the
bare mass. As can be seen in Fig.~\ref{fig:Ev_clov_2635} (bottom right) the
distributions match almost perfectly. The eigenvalue density, for
$|\nu|>0$, on the opposite side of the index peak is almost continuum like
and allows determination of $\hat{m}$ and $\Sigma V$. We use the $|\nu|=1$
eigenvalue distribution for this. The $\hat{a}_j$ are then obtained from
their effect on the index peak. We use the fact that the low-energy
constant have fixed signs $W_6 < 0$, $W_7 < 0$ and $W_8 > 0$
\cite{WRMT,WRMT2,WCHiPT2} and that the distribution depends only on the
combination $|W_6| + |W_7|$ \cite{WRMT} allowing to take $W_7=0$. We find
that with either $\hat{a}_8 \ne 0$ or $\hat{a}_6 \ne 0$ we can reproduce
the histogrammed $|\nu|=0$ and $|\nu|=1$ distributions in
Fig.~\ref{fig:Ev_clov_2635} (top) equally well. But only with $\hat{a}_6
\ne 0$ can we reproduce the $|\nu|=2$ distribution, too, as shown in
Fig.~\ref{fig:Ev_clov_2635} (bottom left).

We can explain the drastically different effect of $W_6$ and $W_8$ on the
analytic prediction for $|\nu|=2$ by noting that the $W_6$-term, in WRMT,
corresponds to a Gaussian fluctuating mass, see Eq.~(\ref{eq:Zfull_WRMT}).
The $\delta$-function index peak of the continuum theory is therefore
smeared into a Gaussian peak with an amplitude that increases with $|\nu|$.
$W_6$, therefore, does not introduce a repulsion between eigenvalues.  On
the contrary, the $W_8$-term of \wchpt\ is included in the representation
of the Dirac operator, Eq.~(\ref{eq:D_WRMT}), of WRMT, and hence induces an
eigenvalue repulsion, as can be seen from the red curve in
Fig.~\ref{fig:Ev_clov_2635} (bottom left). This repulsion is seen for all
sectors with $|\nu|>1$ \cite{WRMT}. It is thus useful to include
eigenvalues from configurations with $|\nu|>1$ for a determination of the
low-energy constants from fits to eigenvalue distributions.

We finally note that, with clover improvement the nonvanishing
$|\hat{a}_6|$ is about a factor 3-4 smaller than the nonvanishing
$|\hat{a}_8|$ without the improvement, both after one HYP smearing,
illustrating again the quite dramatic effect of clover improvement on the
${\cal O}(a^2)$ low-energy constants.

\section{Conclusion}

We have presented numerical simulations, in the quenched case, of the
low-lying eigenvalues of the Hermitian Wilson-Dirac operator, both with and
without clover improvement to compare to predictions from $\epsilon$-regime
Wilson \chpt\ or, equivalently, Wilson RMT. We used the Iwasaki gauge action
which suppresses dislocations and leads to a fairly unique index or
topological charge. This is helpful, since the analytical predictions are
made for sectors of fixed index. We found that our eigenvalue distributions
agree well with the analytical predictions, and verified scaling with
volume and (bare) mass.

We have also looked at the distribution of the real eigenvalues of the
Wilson-Dirac operator, obtained from the spectral flow. We found a dramatic
decrease of both the additive mass renormalization (the real eigenvalue
peak is closer to zero) and the ${\cal O}(a^2)$ low-energy constants (the
width of the distribution becomes narrower and more symmetric) with the
clover improvement.

Fits to the distribution of the low-lying eigenvalues of the Hermitian
Wilson-Dirac operator allow determination of the low-energy constants
of QCD including those that parameterize the ${\cal O}(a^2)$ lattice
effects. However, distributions on configurations with $|\nu|>1$ are
needed to disentangle the effects of $W_8$ from those of $W_6$ and $W_7$
when all $|\hat{a}_j|$ are small.

\section*{Acknowledgments}

UMH thanks the organizers for a stimulating conference and the conveners
of section A for the opportunity to present these results.

\end{document}